\documentclass[a4paper,11pt]{article}
\pdfoutput=1 % if your are submitting a pdflatex (i.e. if you have
\usepackage{jinstpub} % for details on the use of the package, please
\title{\boldmath RPC upgrade project for CMS Phase II}

\author[v,1]{M.I. Pedraza-Morales,\note{Corresponding author.}}
\author[a]{A. Fagot} 
\author[a]{, M. Gul} 
\author[a]{, C. Roskas} 
\author[a]{, M. Tytgat}
\author[a]{, N. Zaganidis}
\author[b]{, S. Fonseca De Souza}
\author[b]{, A. Santoro}
\author[b]{, F. Torres Da Silva De Araujo}
\author[c]{, A. Aleksandrov}
% ...

\affiliation[a]{Ghent university, Dept. of Physics and Astronomy, Proeftuinstraat 86, B-9000 Ghent, Belgium}
\affiliation[b]{ Dep. de Fisica Nuclear e Altas Energias, Instituto de Fisica, Universidade do Estado do Rio de Janeiro, Rua Sao Francisco Xavier, 524, BR - Rio de Janeiro 20559-900, RJ, Brazil}
\affiliation[c]{Bulgarian Academy of Sciences, Inst. for Nucl. Res. and Nucl. Energy, Tzarigradsko shaussee Boulevard 72, BG-1784 Sofia, Bulgaria.}
% ...
\affiliation[v]{Benemerita Universidad Autonoma de Puebla, Puebla, Mexico.}
% ...

\emailAdd{mpedraza@cern.ch}

\abstract{The Muon Upgrade Phase II of the Compact Muon Solenoid (CMS) aims to guarantee the optimal conditions of the present system and extend the $\eta$ coverage to ensure a reliable system for the High Luminosity Large Hadron Collider (HL-LHC) period. The Resistive Plate Chambers (RPCs) system will upgrade the off-detector electronics (called link system) of the chambers currently installed chambers and place improved RPCs (iRPCs) to cover the high pseudo$-$rapidity region, a challenging region for muon reconstruction in terms of background and momentum resolution. In order to find the best option for the iRPCs, an R\&D program for new detectors was performed and real size prototypes have been tested in the Gamma Irradiation Facility (GIF++) at CERN. The results indicated that the technology suitable for the high background conditions is based on High Pressure Laminate (HPL) double-gap RPC. The RPC Upgrade Phase II program is planned to be ready after the Long Shutdown 3 (LS3). 
}

\arxivnumber{1806.11503} % only if you have one

\proceeding{13$^{\text{th}}$ Workshop on Resistive Plate Chambers and Related Detectors\\
  February 19-23, 2018\\
  Puerto Vallarta, Mexico}

\begin{document}
\maketitle
\flushbottom

%\section{LHC Schedule}
%\label{sec:lhc}

\section{Muon Spectrometer}
\label{sec:project}
\begin{figure}[htbp]
\centering % \begin{center}/\end{center} takes some additional vertical space
\includegraphics[width=.9\textwidth]{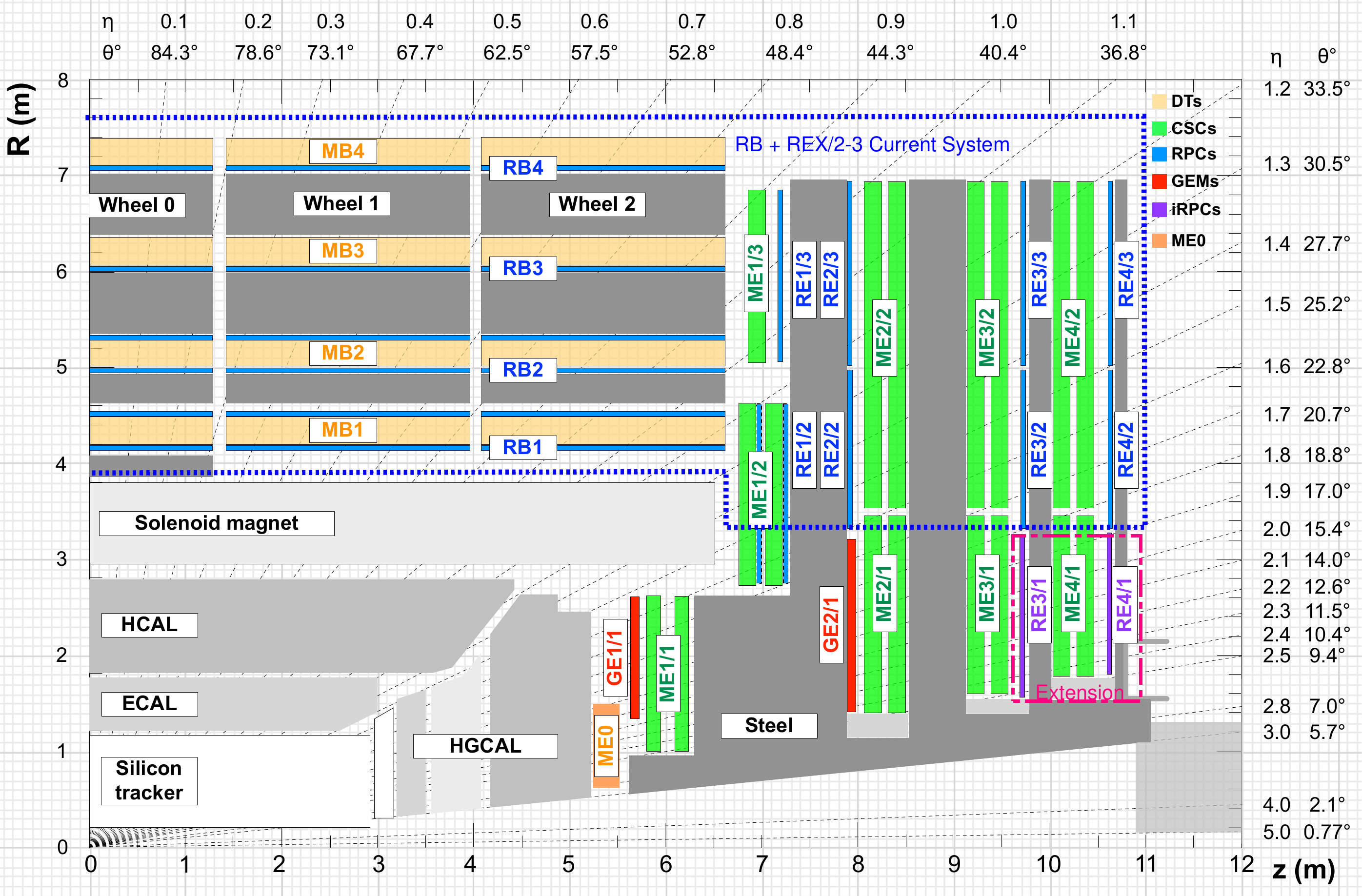}
% "\includegraphics" from the "graphicx" permits to crop (trim+clip)
% and rotate (angle) and image (and much more)
\caption{\label{fig:CMS} An R$-$z cross section of a quadrant of the CMS detector, including the Phase$-$2 upgrades (RE3/1, RE4/1, GE1/1, GE2/1, ME0). The interaction point is at the lower left corner. The acronym iRPCs in the legend refers to the new iRPC chambers RE3/1 and RE4/1. The locations of the various muon stations are shown in color (MB for DTs, ME for CSC, RB and RE for RPC, GE and ME0 for GEM (Gas Electron Multiplier)). M denotes Muon, B stands for Barrel and E for End-cap. The magnet yoke is represented by the dark gray areas.}
\end{figure}

The muon system, described in detail in ~\cite{CMS:1997dma}, uses three different technologies; drift tubes (DT) in the barrel region, cathode strip chambers (CSC) in the end-cap region, and resistive plate chambers (RPC) in both the barrel and end-cap. It covers a pseudo-rapidity region $ |\eta| < 2.4$. The DTs and RPCs in the barrel cover the $\eta$ region $ |\eta| < 1.2$, the CSCs and the RPCs in the end-caps cover the $\eta$ region $0.9 < |\eta| < 1.8$, while the CSCs cover up to $|\eta| < 2.4$. A schematic view of the CMS detector with all Phase$-$2 muon upgrades is shown in Figure \ref{fig:CMS}, where the location of the muon chambers is highlighted. The current RPC system covers $ |\eta| < 1.8$. The barrel region is divided into 5 separate wheels, while the end-caps in 4 disks both in the forward and backward directions, with the 4th disk installed during the first LHC long shutdown in 2013$-$2014 (LS1). In total, there are 1056 RPC chambers, covering an area of about 3950 $m^2$, equipped with about 137,000 readout strips \cite{Pedraza-Morales:2016ysr}.

%The barrel region is divided into 5 separate wheels (named $\pm 2$, $\pm 1$ and 0) while the endcaps in 4 disks both in the forward and backward directions (named $\pm 4$, $\pm 3$, $\pm 2$, $\pm 1$). 

%The 4th disk of the RPCs has been installed during the LHC long shutdown in 2013$-$2014 (LS1) \cite{Mariana}. Each wheel is divided into 12 sectors while every disk into 36 sectors. In total there are 1056 RPC chambers, covering an area of about 3950 $m^2$, equipped with about 137,000 readout strips. The CMS RPCs are double-gap chambers with 2 mm gas width each and copper readout in between. The bakelite bulk resistivity is in the range of $2-5$ $10^{10} \Omega cm$. They operate in avalanche mode with a gas mixture of 95.2\% $C_2H_2F_4$, 4.5\% $iC_4H_{10}$ and 0.3\% $SF_6$.

\section{RPC Muon Upgrade Project}
\label{sec:project}

The LHC has prepared and approved a major upgrade of the accelerators to increase the sensitivity for new physics searches and high precision measurements, the so-called High Luminosity Large Hadron Collider (HL-LHC) period. The instantaneous luminosity will reach a maximum of $7.5 \times 10^{34}$ $cm^{-2}s^{-1}$,  more than seven times the nominal one,  expecting 200 collisions per bunch crossing to collect up to $4000$ $fb^{-1}$ by the end of the LHC operations on 2039~\cite{Collaboration:2283189}. The RPC Muon Upgrade Project consists of the extension of the RPCs system to the very forward region and the sustainability of the current system to ensure a reliable redundancy system for the HL-LHC.  %The link system will be implemented during the Long Shutdown 3 (LS3), while the installation of the $\eta$ extension is expected during the Yearly Technical Stops (YTS) starting at the end of 2021 and 2022. The services related to the $\eta$ extension will be installed during the Long Shutdown 2 (LS2).

\section{$\eta$ coverage extension}
\label{sec:iRPCs}

\begin{figure}[htbp]
\centering % \begin{center}/\end{center} takes some additional vertical space
\includegraphics[width=.55\textwidth]{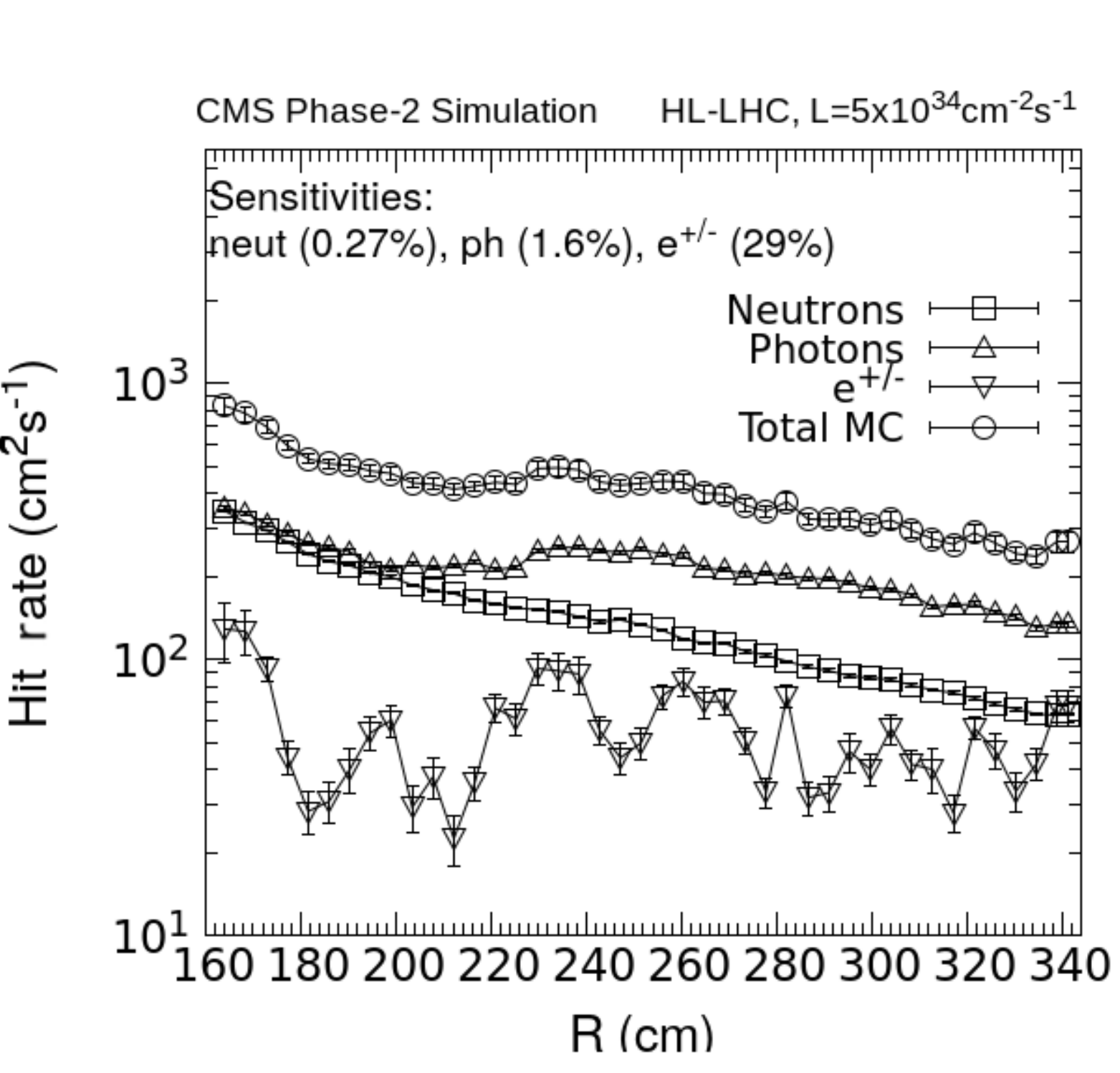}
%\qquad
%\includegraphics[width=.45\textwidth]{Images/eff_vs_eta_station4}
% "\includegraphics" from the "graphicx" permits to crop (trim+clip)
% and rotate (angle) and image (and much more)
\caption{\label{fig:background} Expected hit rate due to neutrons, photons, electrons and positrons at the HL-LHC in the RE3/1 chambers. In the upper part of the figure the sensitivities used in the simulation for each particle are also reported. The expected hit rates in RE4/1 are similar.}
\end{figure}

During the HL-LHC the background induced by neutrons, photons, electrons, positrons and a large number of particles crossing the CMS detector
quasi simultaneously, will reach 700 $Hz/cm^2$ in the hottest points of the region of interest, for an instantaneous luminosity $5 \times 10^{34}$ $cm^{-2}s^{-1}$, figure \ref{fig:background} shows the expected hit rate. This makes the association of individual hits to tracks and thus muon triggering, identification, momentum and charge measurements a major challenge.  To increase the number of hits recorded for a single particle and then improve the muon reconstruction in this difficult zone, the $\eta$ coverage extension is proposed to go from $|\eta| = 1.8$ to $|\eta| = 2.4$. That means adding chambers in the innermost part of the two most distant end-caps of the muon spectrometer, RE3/1, and RE4/1.  

\subsection{Muon identification efficiency}
\label{sec:muonid}

\begin{figure}[htbp]
\centering % \begin{center}/\end{center} takes some additional vertical space
\includegraphics[width=.45\textwidth]{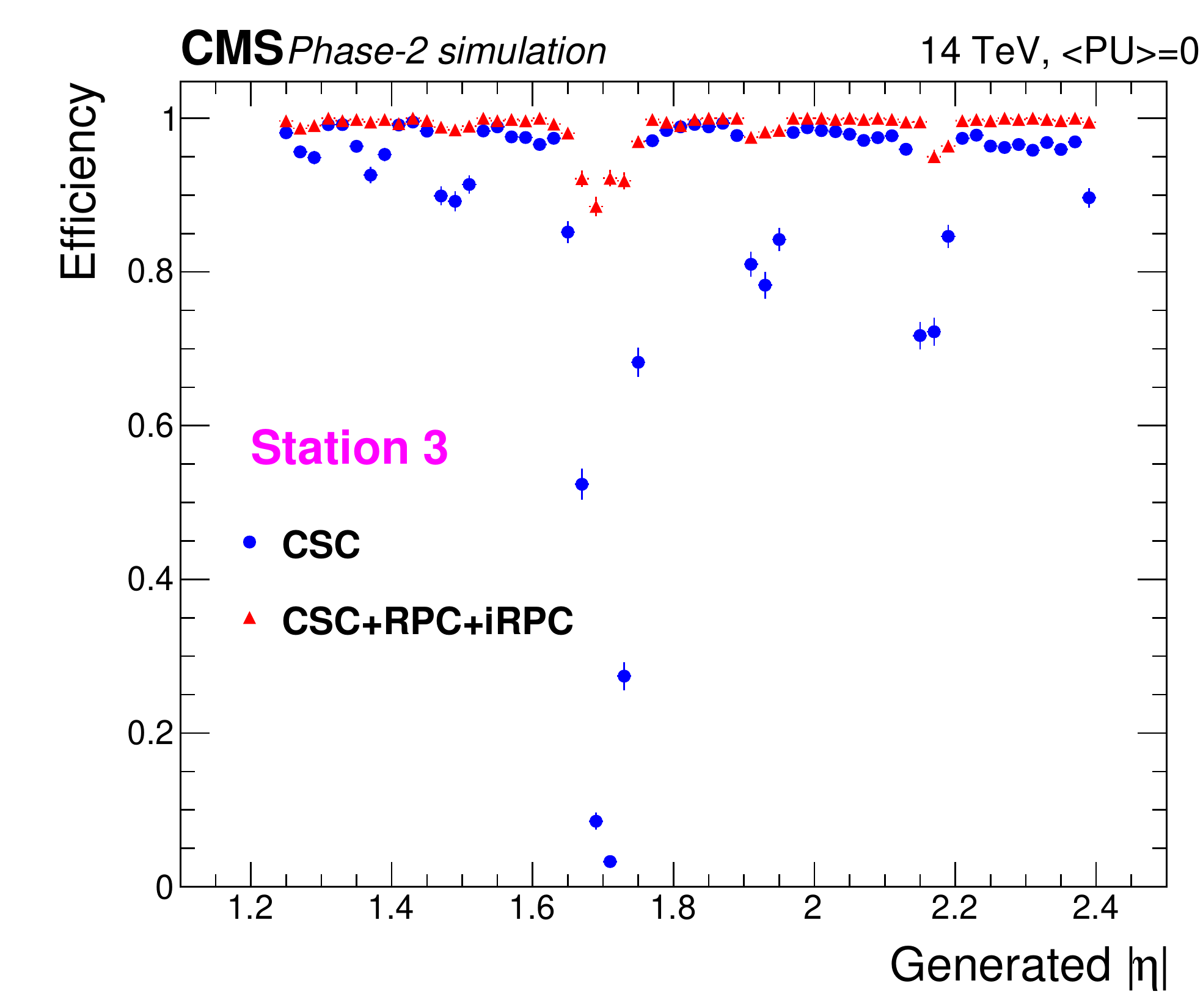}
\qquad
\includegraphics[width=.45\textwidth]{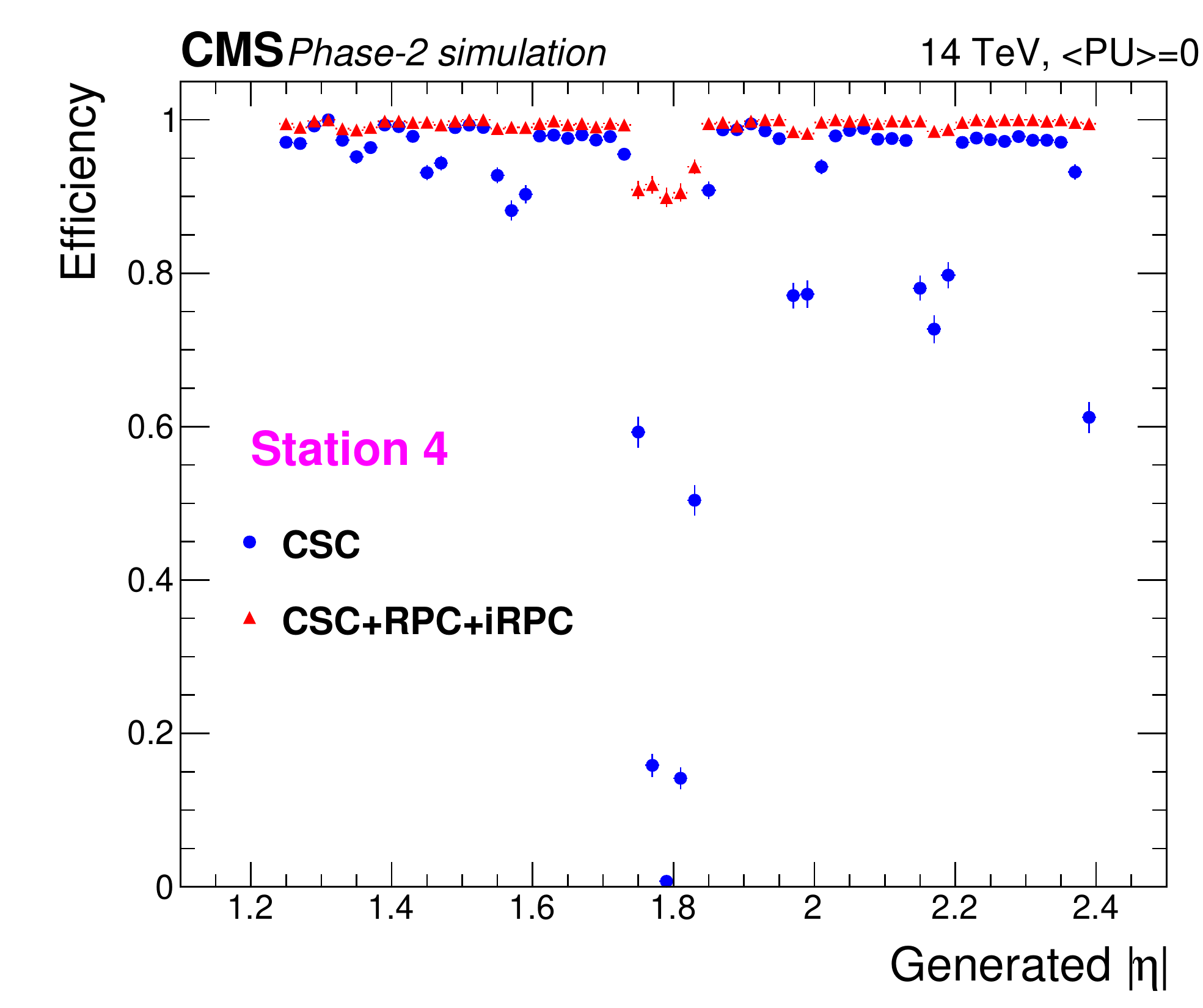}
% "\includegraphics" from the "graphicx" permits to crop (trim+clip)
% and rotate (angle) and image (and much more)
\caption{\label{fig:trigger} Impact of RPC hit inclusion on the local trigger primitive efficiency in station 3 (left) and station 4 (right). The contribution of iRPC starts above $|\eta|$ = 1.8}
\end{figure}

The addition of RE3/1 and RE4/1 will increase the number of hits per muon track. Even though the CSCs system covers the region,  there are steep drops in efficiency in some regions due to the geometry of the system. Including the RPC information into the trigger primitive will mitigate this effect, resulting in an overall increase in efficiency. Figure \ref{fig:trigger} shows the efficiency of finding trigger primitive stubs at the level of station 3 and station 4, with and without the addition of the current RPC and new iRPC information.

%In the present Muon System aging, failures and neutron background simulation

\subsection{Improved RPCs}
\label{sec:improved}

 We have applied a safety factor of three to the hit rate and required that the new chambers stand 2 $kHz/cm^2$ of gamma radiation while maintaining a hit efficiency above 95\%. A double-gap RPC with reduced High Pressure Laminate (HPL) electrode and reduced gas gap thickness, with respect to the present system, has been proposed. These modifications will allow the iRPCs to satisfy the space requirements of the experiment. In addition to satisfy the required performance in rate capability and performance the latest available technology is being used for the design of the Front End Board (FEB) capable of handling the increased background in the forward innermost region. A separate paper has been prepared for this conference.

 %. have been reduced with respect of the design of the current system. It has been shown that reducing the gap thickness can be used to reduce the avalanche charge, and consequently enhance the rate capability \cite{Aielli:2016faq}. 

\begin{figure}[htbp]
\centering % \begin{center}/\end{center} takes some additional vertical space
\includegraphics[width=.5\textwidth]{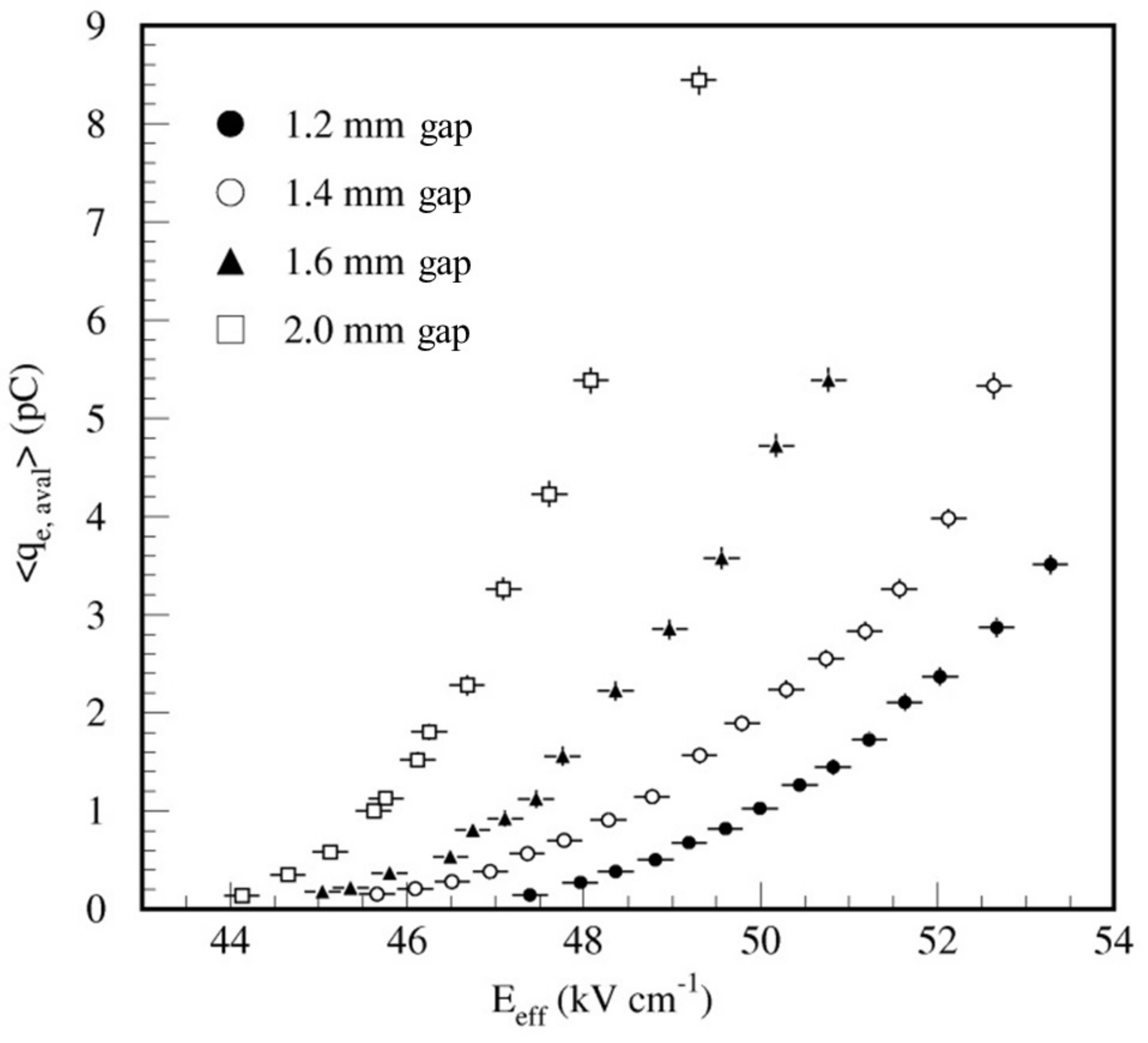}
\caption{\label{fig:iRPCsThickness} Average charge per avalanche measured on 1.20 mm (full circles), 1.40 mm (open circles), 1.60 mm (triangles), and 2.0 mm (squares) double-gap RPCs, as a function of the effective electric field strength.}
\end{figure}

The pickup charge was studied in six double-gap RPCs with gap thickness between 1.0 and 2.0 mm. \cite{Park:2014dxa} at the KOrean DEtector Laboratory (KODEL). The gaps were equipped with a voltage sensitive front-end board with a charge threshold of about 50 fC, which is
lower than the value of the present RPC system front-end electronics by about a factor of three. Figure ~\ref{fig:iRPCsThickness} shows the induced charges drawn in double-gap RPCs for the different gas gap as a function of the effective electric field strength. Electric field values are converted to the effective ones at the reference values of pressure, P = 990 hPa, and temperature, T = 293 K. %Thinner gaps present less average charge per avalanche, which implies that with the correct threshold for the digitization, the operational plateau can be preserved \cite{KSLeeRealSized}. 

This implies that thinner gaps will preserve the operational plateau when the digitization threshold is lowered to increase the detection sensitivity \cite{KSLeeRealSized}. Taking into account that thinner gaps would be more sensitive to non-uniformities, the value 1.4 mm was taken, for the baseline design of the iRPCs, as a safe compromise. 
%The thinner gap thicknesses more effectively retard the fast growth of the pickup charges of the ionization avalanches as shown in Fig. 5.17.
%From TDR: This implies that the use of the thinner gaps will effectively preserve the size of the operational plateau when we lower the digitization threshold to enhance the detection sensitivity \cite{KSLeeRealSized}.

%This solution is the baseline In order to validate We have tested 
%To take into account high eta region fullfill the requirements for in the extended region 

\begin{figure}[htbp]
\centering % \begin{center}/\end{center} takes some additional vertical space
\includegraphics[width=.45\textwidth]{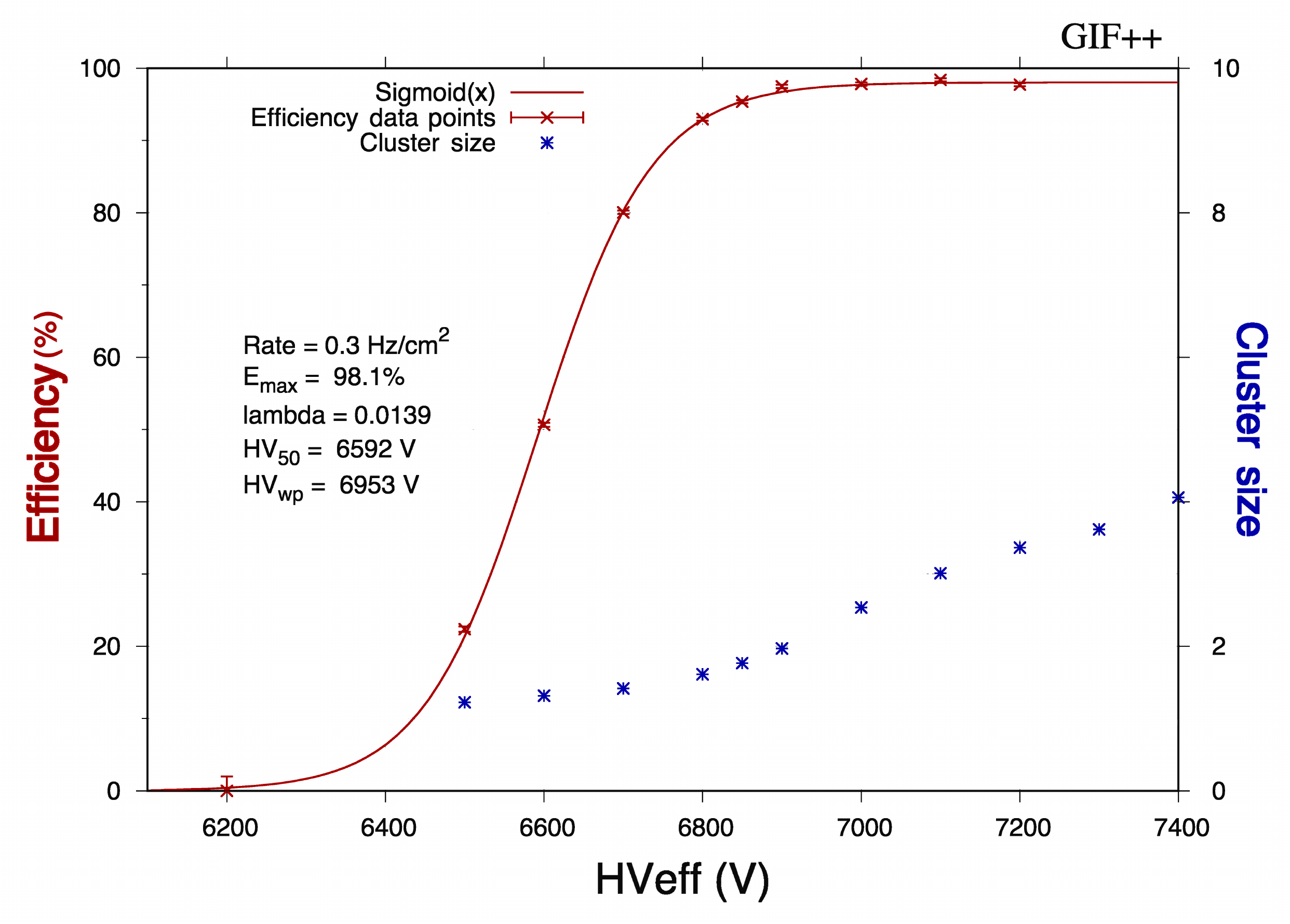}
\qquad
\includegraphics[width=.45\textwidth]{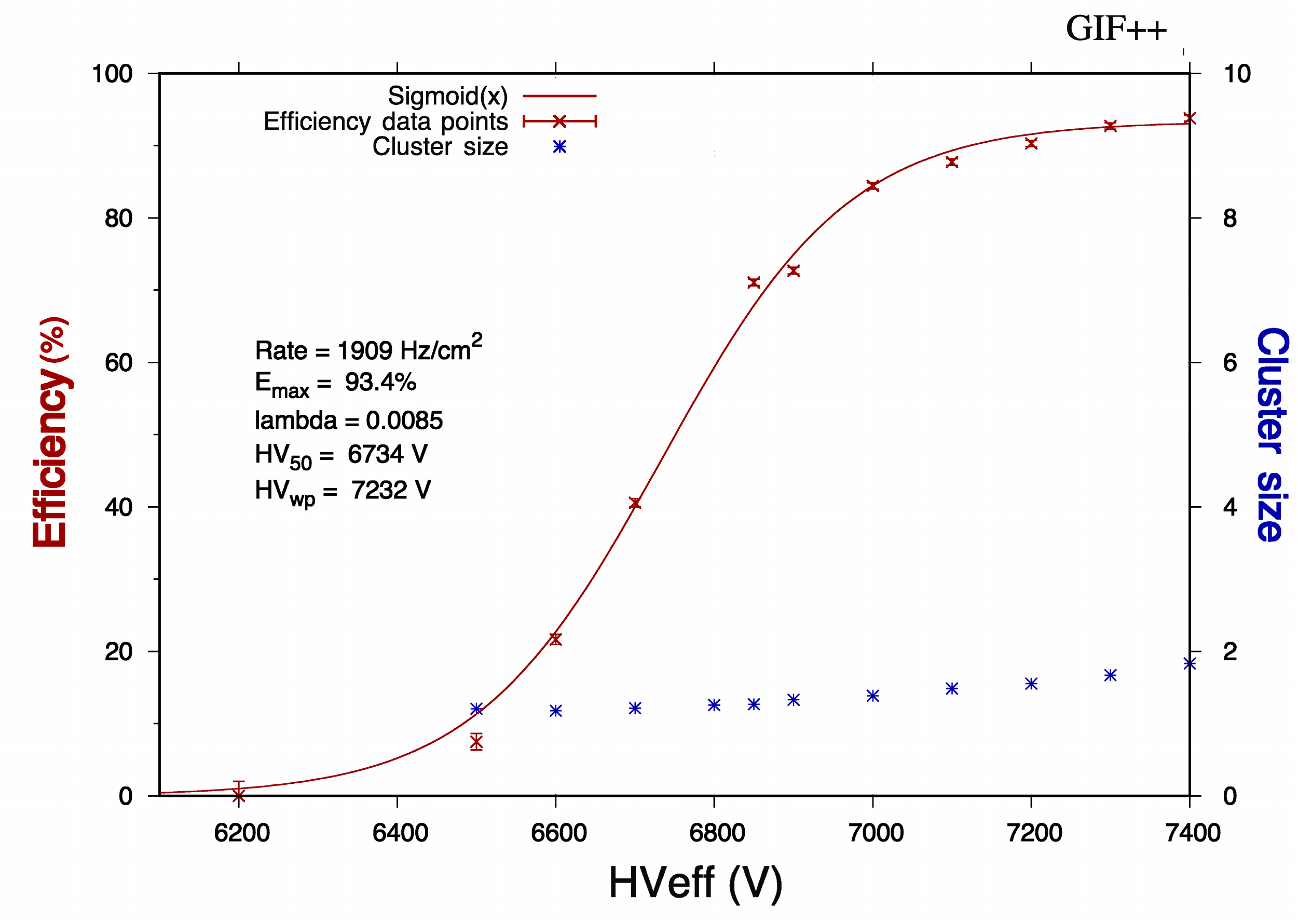}
% "\includegraphics" from the "graphicx" permits to crop (trim+clip)
% and rotate (angle) and image (and much more)
\caption{\label{fig:iRPCefficiency} Efficiency and the average cluster size of a 1.4 mm double-gap RPC large size chamber as a function of the effective voltage, tested without gamma background (left) and under a gamma background rate of 1.91 kHz/cm2 (right). } %The data were measured at the fixed threshold of 300 μV and fitted with a sigmoidal function.}
\end{figure}

In order to validate the performance of the baseline iRPC under different background conditions, a large size prototype was tested at the Gamma Irradiation Facility (GIF++) with muon beam. The prototype was a trapezoidal chamber (long base 92 cm, short base 63 cms, and height 166.3 cm) with 2 cm wide strips and the same front-end electronics threshold used for the tests at KODEL. Figure  ~\ref{fig:iRPCefficiency} shows the rate capability of the prototype with more than 95\% efficiency at $\sim$ 2 $kHz/cm^2$, where the working voltage shift for the highest rate is less than 300 V and a drop on the efficiency of 2\% is observed.

The success of the rate capability test led to the longevity test start and it is in progress at the GIF++ facility. The longevity test aims to secure an optimal performance of the RE3-4/1 after an accummulated charge of 330 $mC/cm^2$, equivalent to more than 10 years of HL-LHC operation. The extended RPC system is expected to enhance the performance and robustness of the muon trigger, corresponding with the coverage of the inner tracker in the extended $\eta$ region, opening the possibility of an independent trigger-based only on the tracker and RPCs hits. 

%The test of the new design including the new electronics has demostrated a good behaivor with no background.

\section{Sustainability of the current system}

The installed CMS RPCs are double-gap chambers with 2 mm gas width that operate in avalanche mode with a gas mixture of 95.2\% $C_2H_2F_4$, 4.5\% $iC_4H_{10}$ and 0.3\% $SF_6$ with a relative humidity of about 40\%-50\%. The link system has the function to process and synchronize the signals coming from the RPC Front-End Boards (FEB) providing RPC hits with a time resolution of one bunch crossing (25 ns). % with 40\% of humidity, working in closed loop mode with about 10\% of fresh gas mixture.
During the HL-LHC operation, the expected conditions in terms of background and pile-up and the probable aging of the present detectors will make the muon identification and correct $p_T$ assignment a challenge for the muon system. In order to make certain the redundancy of the muon system in those conditions, the present RPC system has to be improved.

%each and copper readout in between.  %The bakelite bulk resistivity is in the range of $2-5$ $10^{10} \Omega cm$. 

\subsection{Link Board Upgrade for the current RPC System}
\label{sec:Link}

The Link System of the current RPC, composed of Link Boards (LB) and Control Boards (CB), is responsible for taking the signal from the Front End Boards in the chambers, synchronizing and compressing it, and sending it via the optical links to trigger processors.  In order to secure the good operation of the RPC readout for the HL-LHC there are few weak points that need to be addressed:  

\begin{itemize}
\item Operations: The CBs are connected into token ring logic configuration. Each ring consists of 12 CBs. One ring is responsible for half of a barrel wheel or half of two end-cap disks. If one CB fails then the entire ring does not work, leading to a loss of 6\% of the system.
\item Maintenance: It is a custom electronics, there are not enough LB/CB spares available, and we have to rely on unsupported ASICs.
\item The rate of output optical signals is 1.6 Gb/s. This rate is slow for modern optical receiver interface and optical receivers might not be able to receive data at this rate in the future.
\end{itemize}

\begin{figure}[htbp]
\centering % \begin{center}/\end{center} takes some additional vertical space
\includegraphics[width=.6\textwidth]{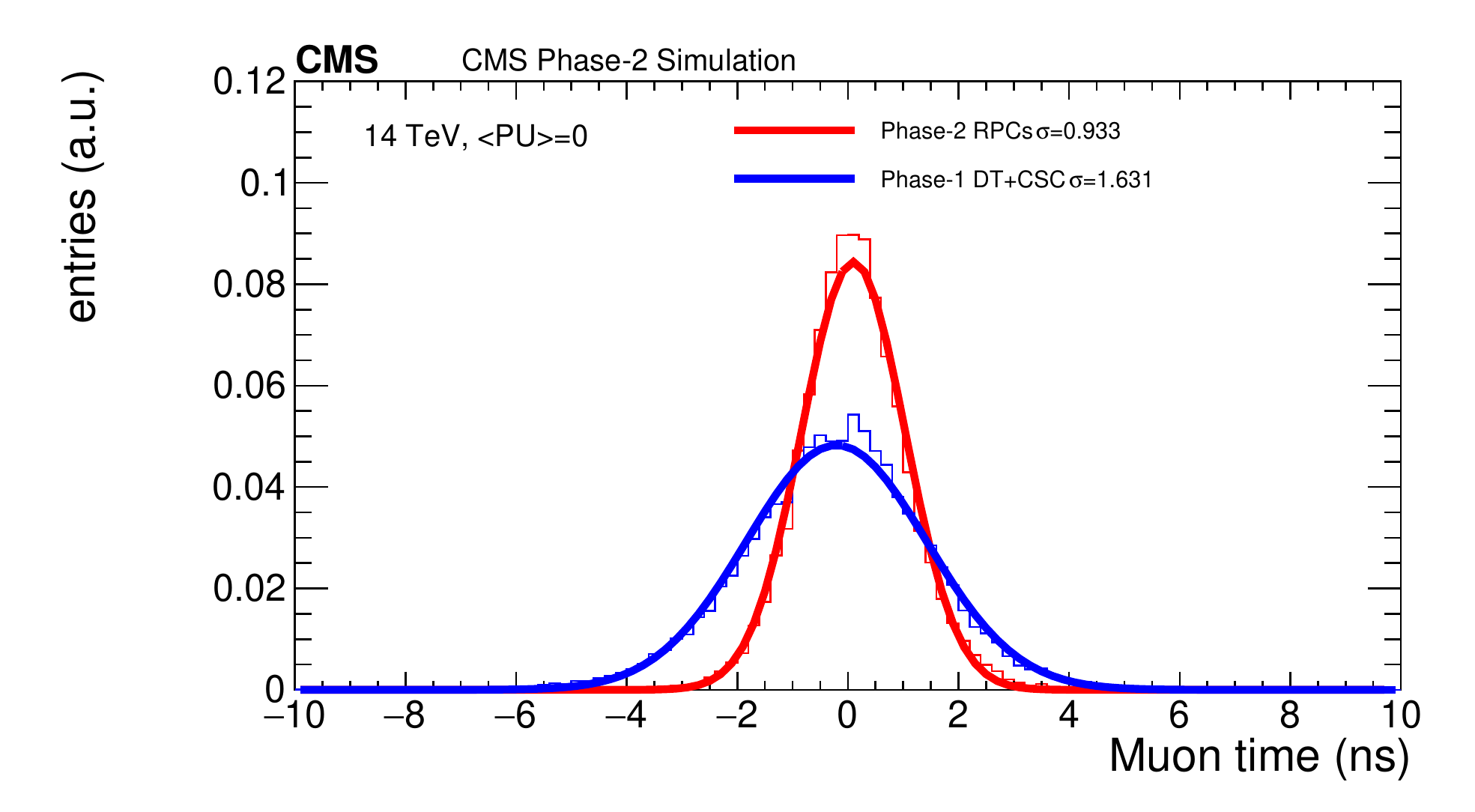}
\caption{\label{fig:LinkSystem} Time residuals between reconstructed and true times associated with a muon for the case when one uses the CSC and DT detectors alone (blue) and for the case of RPCs with the upgraded link system (red).}
\end{figure}

The upgrade of the link system will consist only of the replacement of the LBs and CBs, compatibles with the present Link Boxes, based on Xilinx 7 FPGAs. The new LBs will have the same dimensions, same back-plane connectors, and same pin assignment on the back-plane connectors as in the current system. Current technology allows us to absorb the functionality of most of the chips in the present link system into FPGAs. The ring configuration issue will be resolved by using a high-speed switchboard connecting each CB in parallel. The RPC signal sampling frequency will go from 40 MHz (present system) to 640 MHz (upgraded). Time resolution will be improved from 25 ns to 1.6 ns, having an impact on muon trigger and offline reconstruction. Figure \ref{fig:LinkSystem} shows that the time resolution achievable by using RPCs with the upgraded link system is about a factor two better than the one that the DT and CSC detectors can provide alone, overcoming the weaknesses described before. 

\subsection{Eco-gas studies}
\label{sec:Eco-gas}

The RPC gas mixture mainly consists of fluorinated gases (F-gas) with high Global Warming Potential (GWP) equal to 1307: $C_2H_2F_4$ 95.2\% $SF_6$ 0.3 \% $-$ $iC_4H_{10}$ 4.5\%. Where $C_2H_2F_4$ and $SF_6$ have the highest GWP and are crucial to ensure a stable detector performance in avalanche mode at the background conditions expected at HL-LHC (efficiency more than 95\% and time resolution $\equiv 1.5$ ~ns with a rate capability of $\equiv 600$ $Hz/cm^2$). Recent European regulations on environmental safety policy may lead to a restriction or banning on the use of fluorinated Greenhouse Gases (GHG)*. This has motivated an extensive Research and Development (R\&D) program within the RPC community, two gases with low GWP: HFO-1234ze and $CF_3I$ have been studied previously \cite{Abbrescia:2016xdh} showing promising results. These two gasses have been tested using, in addition, $CO_2$ in order to maintain the HV working point below 12 kV.

\begin{figure}[htbp]
\centering % \begin{center}/\end{center} takes some additional vertical space
\includegraphics[width=.45\textwidth]{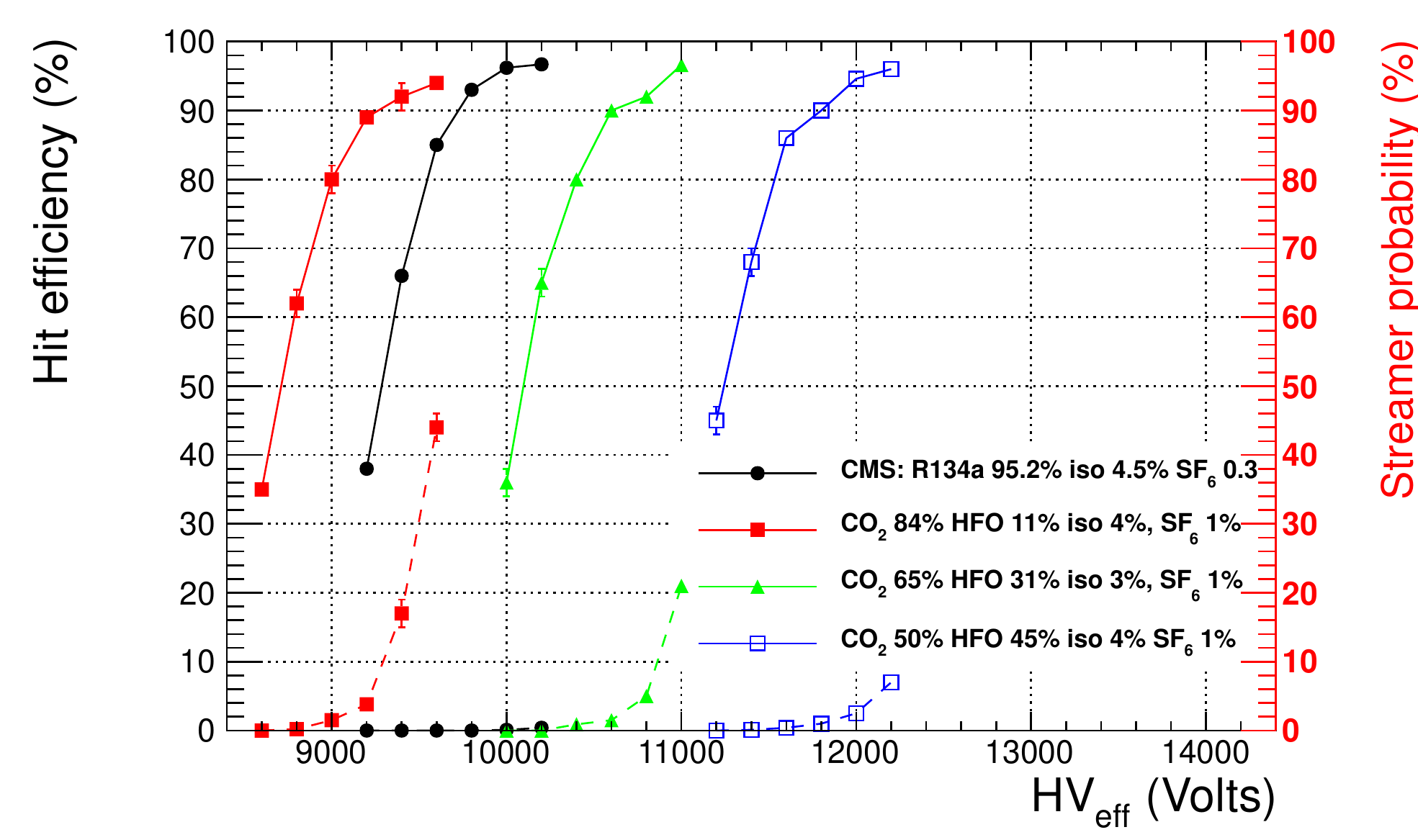}
\qquad
\includegraphics[width=.47\textwidth, height=4.3cm]{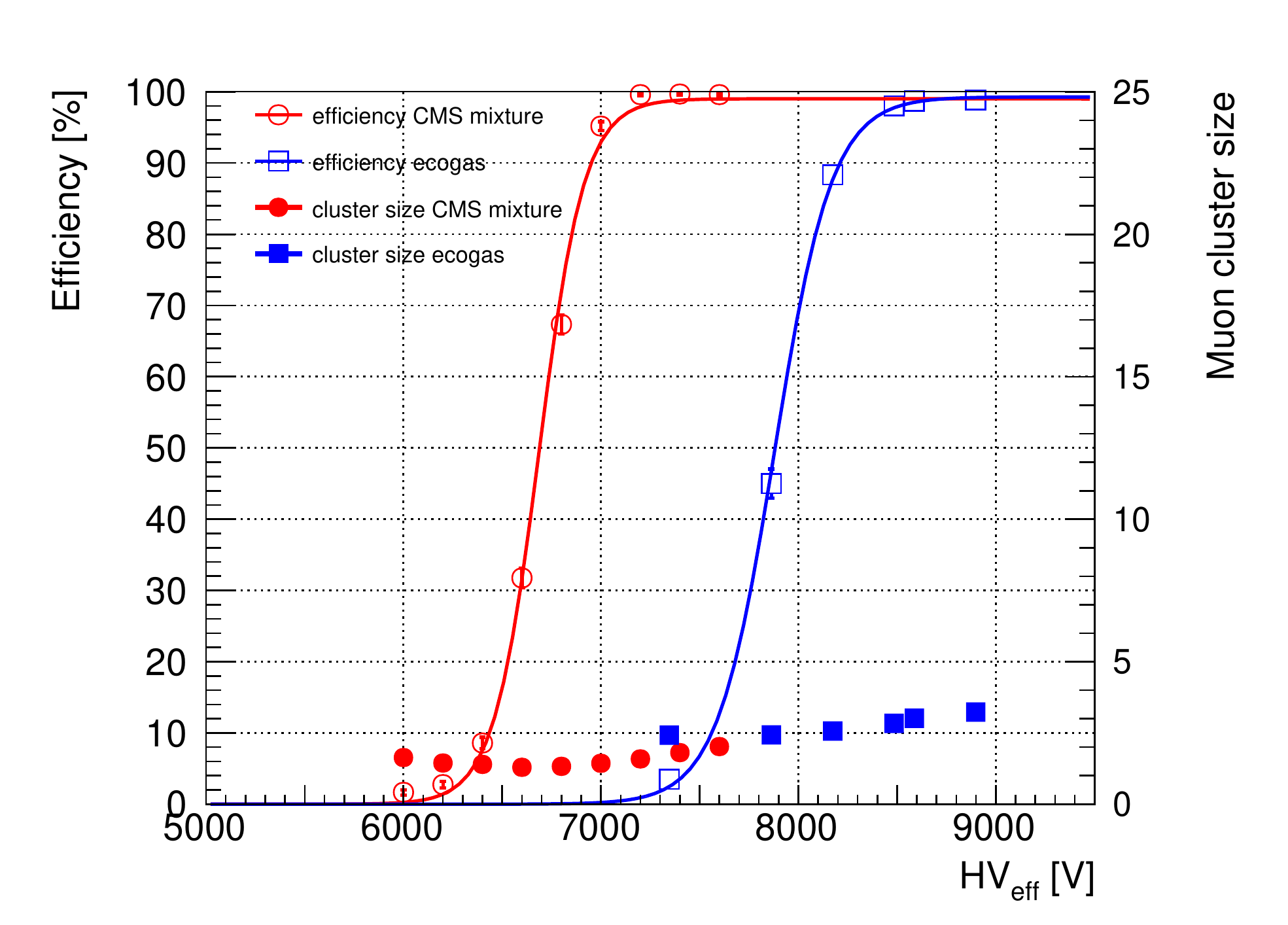}
% "\includegraphics" from the "graphicx" permits to crop (trim+clip)
% and rotate (angle) and image (and much more)
\caption{\label{fig:eco-gas} Hit Efficiency and streamer probability vs. effective voltage for different ecological gas mixtures on RPCs (left). Efficiency and Cluster size vs. effective voltage for the CMS gas mixture and an ecological gas mixture (right) used on iRPCs.}
\end{figure}

Figure \ref{fig:eco-gas} on the left shows that increasing the fraction of HFO$-$1234ze moves the working voltage to higher values and reduces the streamer probability\footnote{Defined as the fraction of events with an induced charge above 20 pC at the working point} at the working point, while on the right % we see the efficiency and cluster size for the iRPCs with 50\% of HFO, showing a 1.3 $kV$ shift in the HV working point. 
we see promising results with 50\% of HFO, showing a 1.3 $kV$ shift in the HV working point. We have seen this with both current and iRPCs, with 50\% HFO, the efficiency is consistent with the one for the standard mixture with up to a 2 $kV$ shift of the HV working point.

%shows the efficiency and cluster size for the iRPCs (left) with 50\% of HFO, showing a 1.3 $kV$ shift in the HV working point, and the efficiency and streamer probability\footnote{Defined as the fraction of events with an induced charge above 20 pC at the working point} for RPC (right) with 45\% of HFO.  

\section{Summary}
\label{sec:Summary}
During the HL-LHC period, triggering on muons, muon identification and correct $p_T$ assignment will be a challenge for the muon system. In order to handle the high background conditions, the number of hits per track should be increased to strengthen the redundancy of the muon system to obtain a solid muon reconstruction. %, improving the rejection of wrongly reconstructed tracks, as early as the Level-1 trigger.  
The required upgrade for the RPC system was presented. It implies to upgrade the link system and to extend the $\eta$ coverage with the installation of iRPCs in the innermost rings of stations 3 and 4. The new link system will improve the precision in the timing readout allowing taking full advantage of the RPC intrinsic time resolution ($\sim$ 1.5 ns), as it recovers the weaknesses and deterioration due to the aging of the present system. And an extensive R\&D program is being performed in order to develop an iRPC that fulfills the CMS requirements for HL-LHC. In this report, we presented the validation of a double-gap large size prototype with a $1.4$ mm electrode and $1.4$ mm gap thickness working with a 95\% efficiency at a rate of 2 $kHz/cm^2$. We also presented promising results for the eco-gas increasing the amount of HFO over the $CO_2$. Both upgrades are expected to be ready after the LHC Long Shutdown 3.
%with an efficiency of 95$\%$.   
%presented the results conserning the performance of different gap sizes and large prot
%This will lead to an increase in the efficiency for both trigger and offline reconstruction in a region where the background is the highest and the magnetic field is the lowest within the muon system. 

%\acknowledgments

%This is the most common positions for acknowledgments. A macro is available to maintain the same layout and spelling of the heading.

%\paragraph{Note added.} This is also a good position for notes added after the paper has been written.


\end{document}